# Underwater Image Enhancement using Generative Adversarial Networks: A Survey

Kancharagunta Kishan Babu[a,*], Ashreen Tabassum[a], Bommakanti Navaneeth[a], Tenneti Jahnavi[a], Yenka Akshaya[a]

[a]*Department of Computer Science & Engineering (AIML & IoT)*
*Vallurupalli Nageswara Rao Vignana Jyothi Institute of Engineering & Technology, Hyderabad-500090, Telanagana, India.*

**Abstract**

In recent years, there has been a surge of research focused on underwater image enhancement using Generative Adversarial Networks (GANs), driven by the need to overcome the challenges posed by underwater environments. Issues such as light attenuation, scattering, and color distortion severely degrade the quality of underwater images, limiting their use in critical applications. Generative Adversarial Networks (GANs) have emerged as a powerful tool for enhancing underwater photos due to their ability to learn complex transformations and generate realistic outputs. These advancements have been applied to real-world applications, including marine biology and ecosystem monitoring, coral reef health assessment, underwater archaeology, and autonomous underwater vehicle (AUV) navigation[1]. This paper explores all major approaches to underwater image enhancement, from physical and physics-free models to Convolutional Neural Network (CNN)-based models and state-of-the-art GAN-based methods. It provides a comprehensive analysis of these methods, evaluation metrics, datasets, and loss functions, offering a holistic view of the field. Furthermore, the paper delves into the limitations and challenges faced by current methods, such as generalization issues, high computational demands, and dataset biases, while suggesting potential directions for future research.

## 1. Introduction

Underwater imaging plays a crucial role in the study and conservation of marine ecosystems, revealing important information on biodiversity, environmental changes, and conservation practices. Researchers can identify species, track population trends, and assess the status of marine environments with high-quality imagery. In industrial use cases like surveys under the water or resource investigations, your visualization must be clear and reliable to avoid safety and disruptions. The capability for this progress is to cut back on reliance on effective techniques and benefit the preservation of marine environments is a promising future that is upon us.

There are some formidable challenges in maintaining underwater image quality. The primary challenge is that light absorption depends on both wavelength and depth. Red light is rapidly absorbed in water, commonly within the initial meters, while green and blue wavelengths can penetrate deeper. Due to this selective absorption, images captured underwater lean more toward blue-green color tones and do not reflect the actual colors of the scene. In addition, the scattering of light due to suspended particles and varying water conditions adds additional problems, such as fogginess, loss of contrast, and lack of distinguishing surface features. Such image distortions present significant challenges for Autonomous Underwater Vehicles and Remote Operated Vehicles, which usually rely on visual data to map, monitor, and navigate.

To overcome these limitations, firstly research was conducted using the physical-free models like HE[2][3][4], CLAHE[5][6], White Balance[7], Retinex[8][9][10], ACDC[11], Wavelet[12], Markov Random Fields[13] (MRF). Then the research was extended to physical models such as RGB[14][15][16], HSI-based[17] methods have been developed to improve efficiency in color correction and image restoration. Furthermore, techniques like Underwater Depth Correction[18] and Underwater Light Attenuation Prior[19] focus on compensating and enhancing image clarity and light attenuation. Despite these advances, it remains difficult to fully compensate for environmental distortions caused by elements like suspended particles and varying water conditions. Image enhancement has experienced multiple progress in deep learning, opening new pathways to improve underwater imagery. Some the deep learning CNN methods include UWCNN[20][21][22], UIE-Net[23], WaterNet[24][25][26], UIECˆ2Net[27][28], Ucolor[29]. While these methods enhance the underwater imagery still they lack in consistently generating enhanced results. Then, to address the challenges, various Generative Adversarial Networks have been proposed, such as WaterGAN[30][31], CycleGAN[32][33], ConditionalGAN[34][35][36][37], DenseGAN[38], MEvo-GAN[39] have been introduced. The two networks that form a GAN—the generator and discriminator—are trained at the same time to learn complex nonlinear mappings from distorted

---

*Corresponding author:
Email addresses:* kishan.kancharagunta@gmail.com (Kancharagunta Kishan Babu), ashreen954@gmail.com (Ashreen Tabassum), bommakantinavaneeth@gmail.com (Bommakanti Navaneeth), jahnavitenneti@gmail.com (Tenneti Jahnavi), akshayayenka@gmail.com (Yenka Akshaya)

to high-fidelity images. The generator repeatedly produces enhanced images, and however, the discriminator assesses their realism. Thus, GANs ultimately generate realistic outputs that restore colors, details, and textures with greater integrity than traditional methods could achieve.

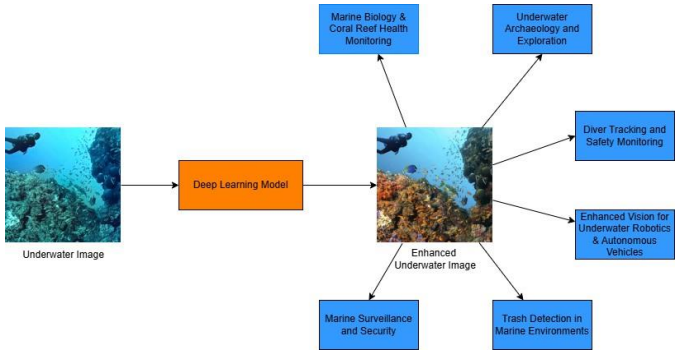

Figure 1: Underwater Image Enhancement Framework

This research is focused on exploring the use of GANs for underwater image enhancement and examining how they can help address many issues seen with traditional methods. By leveraging deep learning-based techniques, this study aims to benefit researchers in developing efficient image enhancement techniques that improve underwater image quality, which has practical applications in both marine research and the industrial sector.

The remaining part of this paper presents a brief understanding of the traditional and deep learning-based enhancement approaches. Then briefing the abilities of GANs and possible uses in the improvement of underwater image recovery. Finally, the research presents a discussion on future directions in underwater imaging and potential improvements in enhancement techniques.

## 2. Literature Review

The literature review is designed with the following criteria that can fit the scope identification of articles and the recent developments in the enhancement of Underwater image(UWI). Following that, the Article selection criteria elaborates on the method of obtaining reliable research papers. Lastly, we focused on the Evolution of UWI Enhancement Techniques which provides significant details on the techniques that have been developed.

### 2.1. Scope of Research

The purpose of this study is to explore numerous techniques that exist to the enhancement of underwater images (UWI) with particular reference to those based on Generative Adversarial Networks (GANs). The research is guided by the following questions:

Q1. What are the various techniques used for UWI Enhancement and how they have evolved?

Q2. In what ways have GAN-based methods improved the enhancement of underwater images as compared with prior methods?

Q3. Which datasets are often used for the assessment of UWI enhancement techniques?

Q4. What are the different performance evaluation metrics implemented in this UWI Enhancement?

Q5. What are the challenges faced while using GANs for underwater image enhancement?

### 2.2. Criteria for Article Selection

In this survey, our primary goal is to review several works regarding underwater image(UWI) enhancement and concentrate on GAN-based techniques in particular. First, we collected the research papers from Google Scholar with keywords relating to "underwater image enhancement", "image deblurring", and "GAN-based methods". This search generated a variety of search strings as follows:

- "UWI" + "enhancement techniques" + "GANs"
- "UWI" + "deblurring" + "GANs".
- GAN + "UWI" + restoration
- "UWI" + available datasets

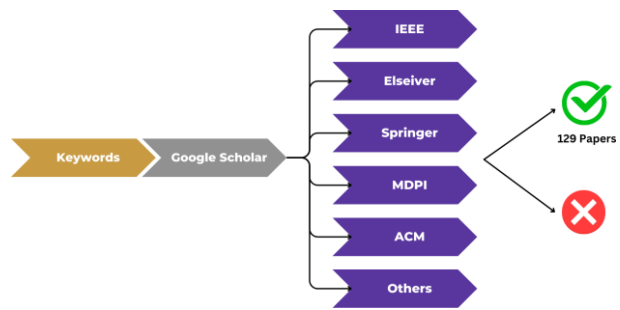

Figure 2: Research Conducted

The findings obtained were further grouped based on the publisher which included IEEE, Elsevier, Springer, MDPI, and ACM as shown in Figure 4. After the categorization step, each paper that was reviewed was assessed on merits as to whether it meets the research interest of the paper at hand which is underwater image enhancement using GANs only papers that specifically fit this research area were considered as outlined in Figure 3. While making the selection, papers that were available in duplicate in an attempt to gain entry into more than one database were also excluded. In addition, only papers with a focus on GAN-based methodology for underwater image enhancement were also not included. Further, specific filters were included based on language, availability of abstracts, and general exclusions where articles with topics such as marine biology that did not have technical or GAN components were removed from the list. The selected papers by category of journals, conferences, and transactions are depicted in Figure 5, and Figure 3 gives the annual trend of the selected topic for underwater image enhancement from 2014 to 2024. The participation of articles has also shown a progressive increase, particularly after the year 2020 and has confirmed the increasing trends of the usage of GANs as well as the deep learning techniques in underwater imaging.

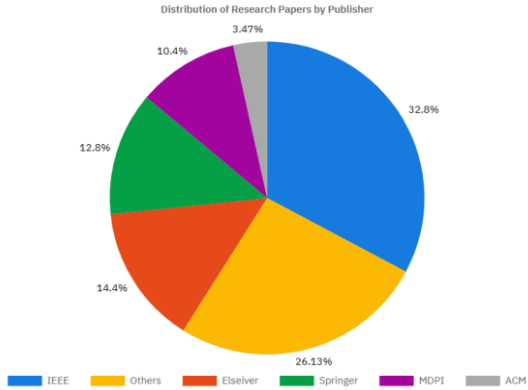

Figure 3: Distribution of Research Papers by Publisher

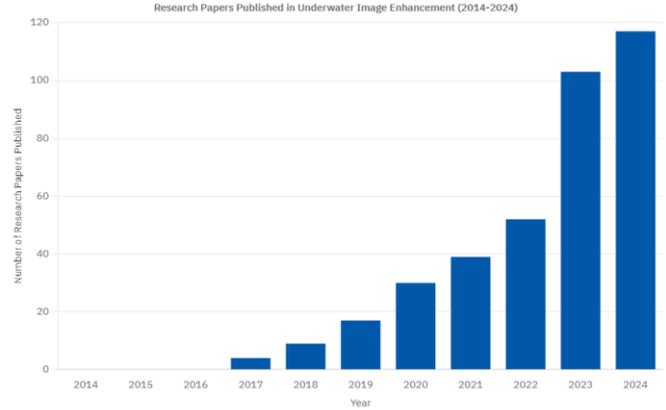

Figure 5: Research Papers Published in a decade

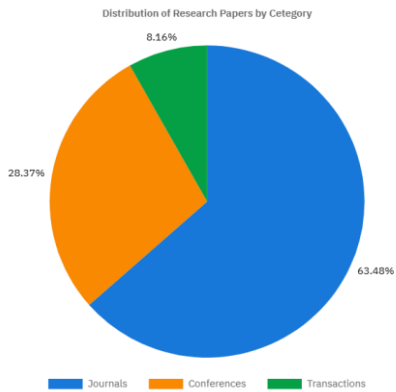

Figure 4: Distribution of Research Papers by Category

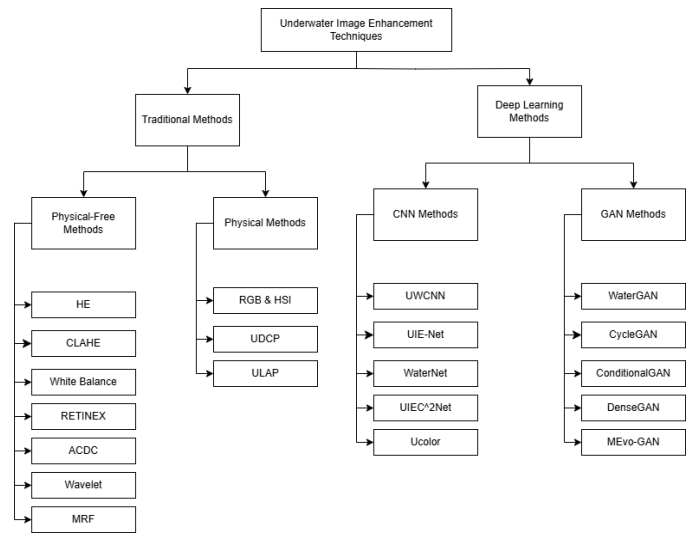

Figure 6: Techniques used in Underwater Image Enhancement

### 2.3. Evolution of UWI Enhancement

In the context of our research work on underwater image(UWI) enhancement techniques, we looked into a number of methods and then categorized them into traditional techniques and deep learning approaches.

On one hand the traditional techniques are grouped into two main categories: Physical-Free and Physical techniques.

Physical-Free Techniques: These are the image quality enhancing methods that do not consider the principles behind underwater imaging such as doesn't rely on scattering, or absorption of light mechanisms. It rather alters the pixel intensities to highlight the contrast and color. Some of the methods are: Histogram Equalization (HE) and CLAHE to enhance contrast White Balance to enhance the precision of color. RETINEX and Wavelet color constancy and detail enhancement.

Physical Methods: Unlike Physical-Free, these methods apply theories of underwater light scatter to address issues such as haze and color cast. Some of them are MRF and RGB & HSI combined for enhanced color representation. UDCP and ULAP use depth cues for increased clarity.

On the other hand research activities in deep learning fall under two broad categories: CNN Methods and GAN Methods.

CNN Techniques: CNNs can learn complex patterns directly from data for quality enhancement. The main Models here are: UWCNN and UIE-Net in rectifying color distortion and haze. WaterNet and UIEC²Net for even more advanced color and detail enhancement to make underwater scenes clearer.

GAN Techniques: The top performance can be achieved by generating images that are actually very real and high-quality
3

underwater. For our study, we researched WaterGAN, and CycleGAN models which allow distorted images to be reformulated into clear, high-resolution views. In total, these methods supply an overarching framework for enhanced underwater image recovery within my project. Traditional methods and deep learning, in particular GAN, present baseline reliable approaches that are limited in some way by creating enhanced high-quality images that enhance underwater visibility and detail effectively.

Quite a few methods have been explored to improve the quality of underwater images such as the classical brightness and contrast enhancement techniques that include Histogram Equalization (HE), CLAHE. As presented in [2], redistribution of pixel intensities makes an effort to lighten up dark regions. However, HE does not rely on some specific aquatic properties of light behavior, which increases noise, color distortion, and irregular results since it increases the contrast all over the image uniformly. Contrast Limited Adaptive Histogram Equalization(CLAHE) [5] is an advanced contrast-enhancement technique to be applied locally on smaller tiles to enhance contrasts. It offers focal enhancements of contrasts by reducing over-exposure effects at the expense of richer details. Standard HE does not consider local variations in intensities and, consequently, is worthless for underwater images whose light attenuation and haze vary with depth. The flexibility of this technique enables it to handle problems such as poor illumination or fog, and increasing color and detail without excessive noise amplification.

White Balance(WB) [7] is another technique that reduces color distortions caused due to various lighting conditions while underwater. The techniques implemented in white balance are Gray-World Assumption and White Patch Retinex, which assume color constancy. It adjusts colors taking into account the average or brightest region. This in turn gives a naturalistic color representation because color casts resulting from varying light being absorbed in water are removed. In this regard, ACDC [11] corrects color images by estimating depth-related color distortion to improve clarity with more specific adjustments.

In addition to color distortion and fuzziness in underwater images, retinex [8]-based methods decompose images into reflectance and illumination components. The Retinex model corrects the color cast and adjusts contrast for better visibility, as it is a balanced enhancement preserving natural image qualities. Layered enhancement, particular to this approach, makes the underwater environment easier in terms of adapting brightness and color while avoiding artifacts.

Wavelet-based methods use the frequency domain approach that involves the application of discrete wavelet transforms (DWT)[12] in decomposing the image into various frequency bands and composing different wavelet layers where color correction and contrast enhancement are applied separately, allowing nuanced synthesis of improved color and detail followed by an inverse transform to reconstruct an image that is visually optimized. The combination of wavelet decomposition with localized adjustments results in better clarity and contrast. However, to bypass the weaknesses of less complex improvement techniques, physical model-based techniques incorporate models that portray underwater physical degradation and algorithms including Dark Channel Prior and Underwater Light Attenuation Prior.

Probabilistic models like Markov Random Fields (MRF) [13] provide learning-based techniques where pixel relationships are modeled statistically. The MRF-based technique imposes color corrections by analyzing pairs of patches with depleted and enhanced colors; hence, it applies all pixel interactions and further color enhancement with a small loss in edges. It enables real-time correction under underwater lighting conditions but is independent of complex physical models.

Besides such methods in image space, other color model alterations, RGB[14] and HSI[17]-based methods adjust color distortion with the uniqueness of color space properties. In RGB-based techniques, light scattering and color absorption are used in a physical model and subsequently undergo inverse processing for correcting an image. Analogously, HSI-based methods adjust saturation and intensity while making use of brightness and color purity to regain the true colors. Such color space conversion allows for sharper images and, therefore, better fidelity in color during color balance correction.

With Underwater Dark Channel Prior(UDCP) and Underwater Light Attenuation Prior(ULAP), the depth-based model that has been developed realizes image enhancement that continues its analysis of effects particular to underwater scene absorption and scattering. UDCP[18] is targeted especially on reversing a model based on light attenuation as compensation for the rapid absorption of red light. On the other hand, ULAP[19] uses depth estimation in deriving images to produce transmission maps that correct hazy and color distortions at multiple levels of depth, thereby providing effective color correction with minimal computational load. The above traditional methods for underwater image enhancement such as HE, CLAHE, and UDCP elevate the contrast and color but neglect the properties of underwater light or incorporate them in the high computational-time model. Other techniques like the CNN models surpass these challenges since they capture the intricate patterns in an image with great success in remedies of color shift and haze.

**CNN METHODS**: Chongyi Li et al.[20] have proposed UWCNN (Underwater Scene Prior-Inspired Convolutional Neural Network) which combines underwater scene prior knowledge and synthetically generated datasets based on NYU-v2 dataset to enhance the performance of a lightweight CNN model for different water types and degradation levels underwater. The architecture includes three enhancement units with convolution layers, ReLU activation function, and residual learning, with ten layers and sixteen feature maps. To assess the performance of the model different parameters such as MSE, PSNR, and SSIM are used which are better than previous techniques used but they face challenges like low dynamic range, and low contrast in real-world underwater scenarios.

Wang et al.[23] proposed UIE-Net, a CNN-based framework used for enhancing underwater images due to color distortion and problems with haze. It employs two subnetworks: a Color Correction Network (CC-Net) for color changes and a Haze Removal Network (HR-Net) for contrast enhancement. The model is trained on 200,000 synthesized images derived from a physical underwater imaging model. However, it heavily relies on synthetic data due to limited labeled real-world underwater datasets and also involves high computational complexity.

The paper proposed by Li et al.[40], presents UIEB, a benchmark for underwater image enhancement, and presents Water-Net, a CNN model developed on the proposed dataset. The UIEB is a dataset of 950 raw underwater images. In particular, Water-Net uses a gated fusion network to effectively enhance image quality depending on the confidence maps where the network uses preprocessing techniques such as white balance, histogram equalization, and gamma correction. However, the approach also presents drawbacks such as the inability to fully eliminate backscatter effects, the fact that the enhancement algorithms applied are built based on inaccurate physical models, and the synthetic training data appropriately generalizable to real-world conditions.

Wang et al.[27] introduced a novel CNN-based framework for underwater image enhancement known as UIEC²-Net that incorporates both the RGB and HSV color channels. The model comprises three main blocks: An RGB pixel-level block for simple tasks like de-noising and color cast elimination, an HSV global-adjust block for fine-tuning the luminance and saturation of enhanced images, and an attention map block that integrates the outputs of the other two blocks to generate high-quality enhanced images. However, some limitations include the sensitivity of the HSV hue channel which could lead to color distortions and the problem of synthesizing training data from real-world scenarios.

Further, Hou et al.[29] presents a method called Ucolor which uses a multi-color space encoder to encode multiple feature representations and uses a channel-attention mechanism to emphasize discriminative features. Moreover, a Medium Transmission Decoder is developed to combat the degraded quality domains. However, it faces challenges due to its reliance on high-quality data used during the training process.

While each of these methods has an added unique advantage in enhancing images underwater, it still presents challenges in the ability to have consistently proper results among different underwater conditions. Under the light of these understanding criteria, GANs can hold great promise by learning adaptively to enhance underwater images based on extensive data without any predefined model for light behavior or manual tuning of parameters.

## 3. Underwater Image Enhancement using GANs

Generative Adversarial Networks (GANs) are a type of neural network architecture used in machine learning to generate new data that resembles a given dataset. They were introduced by Ian Goodfellow[41] in 2014 and have become highly popular for creating realistic images, videos, and other data formats. The unique approach in GANs involves two networks that "compete" against each other, which leads to the generation of high-quality, realistic data.

### 3.1. The GAN Architecture:

The architecture of a Generative Adversarial Network (GAN) consists of two neural networks: the Generator and the Discriminator

**Generator:** The generator is responsible for creating synthetic data samples that closely resemble real data from the training set. It begins with random noise as input and processes this through multiple layers to generate structured, realistic data, like an image or text. The Generator's primary objective is to minimize generator loss, which reflects how well it can "fool" the Discriminator into classifying its outputs as real. As training progresses, the Generator continuously adjusts its parameters to reduce this loss, ultimately improving its ability to produce outputs that closely mimic genuine data.

**Discriminator:** The discriminator in a GAN functions as a binary classifier, distinguishing between real samples from the training data and synthetic samples generated by the Generator. It assigns a probability to each input, indicating its belief that the data is real or fake. The Discriminator's goal is to minimize discriminator loss, which reflects its accuracy in correctly identifying real versus generated data. As it trains, it updates its parameters to improve classification, resulting in decreased Discriminator Loss as its detection accuracy improves. This adversarial dynamic compels the Generator to produce increasingly realistic data, with both networks continuously refining their outputs and classifications to achieve high quality.

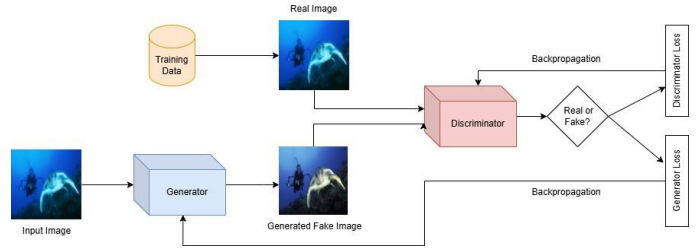

Figure 7: Generative Adversarial Networks Architecture

### 3.2. Underwater Image datasets:

Below, some real-world underwater image datasets are familiarly known in the marine sciences.

**EUVP**: The EUVP[42] (Enhancing Underwater Visual Perception) dataset includes three paired subsets: Underwater Dark (5,550 training pairs, 11,670 total images), Underwater ImageNet (3,700 training pairs, 8,670 total images), and Underwater Scenes (2,185 training pairs, 4,500 total images). The unpaired dataset contains 3,195 poor-quality images and 3,140 good-quality images, with a total of 6,665 images[43].

**UIEBD**: UIEBD[40] (Underwater Image Enhancement Benchmark Dataset) comprises 950 images underwater. and 890 high-quality reference images derived from different enhancement methods. It includes two subsets: one with 890 normal images paired along with high-quality images and another with 60 challenging samples.

**Fish4Knowledge**: The Fish4Knowledge[44] dataset, funded by the European Union's Seventh Framework Program, is a substantial resource designed for studying marine ecosystems. It comprises video and fish analysis data, totaling around 200 TB in size.

**ULFID**: The ULFID[45] (Underwater Light Field Image Dataset) consists of light field images captured in both clear and hazy underwater conditions, along with corresponding in-air reference images. This dataset allows researchers to evaluate image enhancement techniques under various lighting conditions, facilitating advancements in underwater imaging and analysis.

**MARIS**: MARIS[46] (Marine Autonomous Robotics for InterventionS) focuses on advancing the development of cooperating Autonomous Underwater Vehicles (AUVs) for various tasks which include search-and-rescue operations and scientific exploration. The project captures underwater images and videos using stereo vision systems, enhancing the capabilities of robotics in undersea environments and contributing to various intervention efforts in the offshore industry.

**SQUID**: SQUID[47] is comprised of 57 stereo pairs acquired from four dive sites of Israel, including two sites in the Red Sea—'Katzaa' (a coral reef at depths of 10-15 meters) with 15 pairs and 'Satil' (a shipwreck at 20-30 meters) with 8 pairs—and two Mediterranean rocky reef sites: 'Nachsholim' (3-6 meters deep) with 13 pairs and 'Mikhmoret' (10-12 meters deep) with 21 pairs. The dataset includes various image formats, such as RAW images and TIF files, along with camera calibration files and distance maps, totaling around 33GB in size.

**RUIE**: The RUIE[48] dataset addresses underwater image enhancement challenges by providing a benchmark to evaluate visibility degradation, color casts, and detection accuracy. Captured with a multi-view system of twenty-two waterproof cameras near Zhangzi Island, it includes about 4,000 images taken at depths of 0.5 to 8 meters. The dataset features three subsets: the Underwater Image Quality Set (3,630 images for visibility improvements), the Underwater Color Cast Set (300 images for color correction), and the Underwater Higher-level Task-driven Set (300 labeled images for testing classification effectiveness).

**U-45**: The U45[49] dataset for instance is made of 45 real under sets of water images assembled to experiment with image improvement techniques. It is derived from 6,128 image pairs which have been obtained from the underwater GAN and other origins to fill in an evident gap in the current methodologies. These images are organized into three categories: Green, Blue, and foggy appearance—proving different sorts of degradation like color, casts, and low contrast. This variety allows the testing of enhancement algorithms for low and high-level video processing application areas.

*3.3. Evaluation Metrics:*

**MSE**: Mean Squared Error (MSE) is an evaluation parameter that assesses the disparity between two signals, commonly the original signal and its version is either distorted or enhanced. It measures the square of the differences in the pixels involved and then finds the mean of all of these. in the two images giving an average measure of enhancement error[50]. The formula for MSE is given by:

$$MSE = \frac{1}{S} \sum_{j=1}^{S} (U_j - Ue_j)^2$$

$U_j$ indicates actual value and $Ue_j$ indicates predicted value.

**PSNR**: Peak Signal-to-Noise Ratio (PSNR) measures the quality of the image. "Clarity as represented by an improved picture concerning a basic picture," that is. the extra noise that creeps into the system during enhancement. Higher PSNR values imply improved image quality since they signify that the enhancement process has helped to reduce noise while at the same time enhancing important details[51]. This is because PSNR is determined from the Mean Square Error(MSE) using the formula:

$$PSNR = 10 \log_{10} \frac{L^2}{MSE}$$

L represents the dynamic range of pixel intensities (typically 255 for images).

**SSIM:** The Structural Similarity Index (SSIM) evaluates the quality of images by comparing luminance, contrast, and structure, mirroring human visual perception. It assesses how well-corrected images maintain natural details relative to ground-truth images. Higher SSIM values indicate better preservation of essential features in underwater scenes. The local SSIM can be computed using the formula:

$$SSIM(u, v) = \frac{(2\mu_u\mu_v + k_1)(2\sigma_{uv} + k_2)}{(\mu_u^2 + \mu_v^2 + k_1)(\sigma_u^2 + \sigma_v^2 + k_2)}$$

where the average luminance of the patches is represented by $\mu_u$ and $\mu_v$, while $\sigma_u^2$ and $\sigma_v^2$ denote the variances. The covariance is indicated by $\sigma_{uv}$, and $k_1$ and $k_2$ are constants included to prevent instability in the division process.

**UCIQE:** UCIQE assesses color balance in underwater images by measuring chroma, contrast, and saturation brightness. The expression for UCIQE is given as:

$$UCIQE = k_1 \cdot \sigma_c + k_2 \cdot \mathbf{lcon} + k_3 \cdot \mu_s$$

where $\sigma_c$ represents standard deviation of chroma; lcon is the luminance contrast, indicating brightness differences; and $\mu_s$ is the mean saturation, measuring color intensity. Constants $k_1$, $k_2$, and $k_3$ are set to weigh each component, making UCIQE suitable for assessing underwater image quality where color fidelity and brightness are affected by water conditions.

**UIQM:** Underwater Image Quality Measure (UIQM) assesses image quality by integrating three attributes: colorfulness (UICM), sharpness (UISM), and contrast (UIConM). UIQM does not require a reference image, making it ideal for practical underwater conditions where comparison images aren't available. The formula for UIQM is given as:

$$UIQM = k_1 \times UICM + k_2 \times UISM + k_3 \times UIConM$$

where $k_1$, $k_2$, and $k_3$ are application-specific weights that emphasize different aspects of image quality based on the needs of the underwater environment.

Table 1: Literature Review

| Paper Reference | Dataset | Availability | Data Observations | URL |
|---|---|---|---|---|
| [42] | EUVP | Public | 31,505 Images | link |
| [40] | UIEBD | Non-commercial, Academic | 950 (890 with references, 60 challenging) Images | link |
| [44] | Fish4Knowledge | Private | 4 million fish observations (video data) | - |
| [45] | ULFD | Public | Images taken from pure and Muky water | link |
| [46] | MARIS | Public | 10,123 stereo images | link |
| [47] | SQUID | Public | The dataset contains 57 stereo pairs from four different sites in Israel (2 in the Red Sea, 2 in the Mediterranean Sea) | link |
| [48] | RUIE | Public | Three subsets: UIQS- 3,630 images UCCS- 300 images UHTS- 300 images | link |
| [49] | U-45 | Public | Three subsets: green, blue, and haze | link |

**PCQI:** The Patch based Contrast Quality Index (PCQI) is an IMQ that can be used to assess local image quality for image databases. It presents a method for computing the contrast in an image in comparison to a patch-based approach. However, in traditional methods that estimate global image statistics, PCQI focuses on local image patches. This index assesses three key components in each patch: signal mean level, signal power, and structure which permits a further analysis of image quality at a local level. The expression for PCQI is given as:

$$PCQI = q_{mi}(u, v)q_{sd}(u, v)q_{cc}(u, v)$$

$q_{mi}(u, v)$ indicates mean intensity, $q_{sd}(u, v)$ gives structural distortion, and $q_{cc}(u, v)$ determines the contrast change.

*3.4. GAN based Methods*

In this section, we have provided a state-of-art review of the research work which is carried out in context to Underwater Image (UWI) enhancement, particularly the GAN-based approach.

*3.4.1. WaterGAN*

Though both WaterGAN and Fusion Water-GAN (FW-GAN) are designed to specifically solve the problems associated with underwater image degradation they use different principles and architectures adjusted for different purposes. The WaterGAN was proposed by Li et al. (2017) [30] where they incorporated a GAN pipeline for generating and correcting underwater color images. WaterGAN synthesizes underwater images from in-air RGB-D pairs by modeling the light propagation through the water as well as capturing interactions such as absorption, scattering, and geometric distortions caused by the camera. The generator architecture in WaterGAN is presented in three stages:

- Attenuation Stage: Performs range-dependent fading of light following the simplified Jaffe-McGlamery model to adapt each of the RGB channels to the wavelength.

- Scattering Stage: Uses the back-scattered light to make the haze effect that could be seen in the underwater pictures.

- Camera Model Stage: Mimics real-world camera effects, such as vignetting and a linear sensor response, controlling the brightness with a gradient in the vignetting mask.

The discriminator in WaterGAN is a convolutional network designed to differentiate between real underwater images and synthetic ones generated by the model. Through convolutional layers and leaky ReLUs for nonlinearity, the discriminator classifies images as real (1) or synthetic (0), encouraging the generator to produce highly realistic underwater images. Water-GAN is further enhanced by a color correction network that performs real-time color restoration on actual monocular underwater images, incorporating depth estimation and skip connections within a fully convolutional encoder-decoder structure. Validated on datasets from controlled tank tests and field surveys, WaterGAN excels in creating realistic underwater imagery with improved color consistency and accuracy, which has applications in 3D mapping and visual clarity enhancement for underwater robotics. However, WaterGAN's assumptions about lighting and centered vignetting patterns pose limitations, as the model often requires site-specific retraining to adapt to varied environmental conditions and lighting complexities.

In contrast, FW-GAN, a multi-scale fusion GAN architecture proposed by Wu et al.[52], directly enhances real underwater images, focusing on improving visual quality in degraded underwater scenes by presenting common issues such as color distortion, low contrast, and haze. Unlike WaterGAN's synthetic image generation, FW-GAN is designed specifically to refine real-world underwater images using a U-Net-based generator with an encoder-decoder structure. To adapt to underwater conditions, FW-GAN fuses multiple enhancement priors, including White Balance (WB), Contrast Limited Adaptive Histogram Equalization (CLAHE), and Dark Channel Prior (DCP). These prior features are encoded, refined, and fused at various scales within the network, passing through a channel attention decoder that emphasizes crucial feature channels to generate high-quality output. The FW-GAN discriminator, based on a PatchGAN structure, retains high-frequency details, improving generalization for different underwater settings and achieving stable training with spectral normalization. FW-GAN was validated on both synthetic (EUVP) and real-world (UIEB) datasets, showing higher scores on image quality metrics such as PSNR, SSIM, UCIQE, and UIQM. Its strength lies in enhancing underwater image quality across varied environments, aiding tasks like feature recognition and saliency detection for underwater robotic vision. Nonetheless, FW-GAN's computational intensity and occasional limitations in extreme underwa-ter conditions highlight potential areas for optimization, such as improving efficiency and robustness under varied lighting and color distortions.

Though both methods use GAN-based approaches for underwater image enhancement, WaterGAN and FW-GAN differ in their goals and execution. WaterGAN focuses on generating synthetic datasets that simulate underwater effects for applications in mapping and visual reconstruction, requiring precise environmental data and often site-specific retraining. FW-GAN, however, is designed for direct enhancement of real underwater images, offering a generalized solution across different environments by integrating multi-scale fusion and channel attention mechanisms. This difference underscores the unique contributions of each framework: WaterGAN provides a scalable solution for creating synthetic datasets under controlled assumptions, whereas FW-GAN enhances practical visual quality for immediate applications in underwater robotics and inspection tasks. Addressing their respective limitations could involve enhancing WaterGAN's adaptability to diverse lighting conditions and refining FW-GAN's computational efficiency to allow real-time processing on resource-limited hardware.

*3.4.2. CycleGAN*

Several studies have explored the use of CycleGAN, an unpaired image-to-image translation framework, to present the unique challenges of underwater image enhancement, particularly issues related to color distortion, light scattering, and noise in underwater environments [53, 54, 32]. CycleGAN, introduced by Zhu et al. (2017)[32], provides a robust architecture for translating images between two domains without requiring paired data, a crucial feature for underwater image processing where paired datasets are often unavailable[32]. This framework uses a cycle-consistency loss, which ensures that an image transformed from one domain to the other and back again approximates the original, thus facilitating high-fidelity domain adaptation for underwater imagery [32].

The study by Fabbri, Islam, and Sattar (2018) applies CycleGAN to underwater images by transforming distorted visuals into clearer representations similar to above-water images [53]. This enhancement addresses visual challenges faced by autonomous underwater vehicles (AUVs), particularly in tasks like diver tracking and seabed mapping, where color accuracy and image clarity are essential. Fabbri et al. utilize a U-Net-based encoder-decoder model as the generator, equipped with skip connections that preserve spatial information, helping to capture detailed textures and overall scene structure. Alongside this generator, they employ a PatchGAN discriminator that processes small image patches (e.g., 70x70 pixels), allowing it to detect local inconsistencies and restore image fidelity by focusing on fine details. This architecture enables the model to restore color and reduce noise in underwater images by learning complex transformations between "clear" and "distorted" underwater images. The study evaluates the model's performance through metrics like edge detection using the Canny method and Gradient Difference Loss (GDL) to assess image sharpness and color consistency. They also measure the practical improvement of AUV tracking accuracy, finding that the model's transformations enhance the performance of diver tracking using the Mixed Domain Periodic Motion (MDPM) tracker. These findings underscore CycleGAN's potential to significantly enhance image clarity in underwater environments, though performance can still vary based on specific underwater conditions [53].

Another adaptation, UWGAN, applies CycleGAN principles specifically for color correction in underwater images, focusing on weakly supervised training to improve image quality without requiring paired data [54]. UWGAN introduces a CycleGAN-based structure that utilizes adversarial, cycle-consistency, and Structural Similarity Index Measure (SSIM) losses to achieve realistic color adaptation while preserving important structural details. The adversarial loss assist the model in learning how to produce air-like images from underwater inputs and the cycle consistency loss increases high-fidelity transformations, therefore consistency loss benefits are apparent. The addition of SSIM loss is to make sure that the model learns the structural integrity suitable for detailed applications including intensive tasks, like saliency detection for image preservation or keypoint matching. Leaning from 3,800 images of underwater environment and 3,800 of air images, UWGAN successfully increases the accuracy of color reconstruction. Though as it has the capability of performing differently depending with the various underwater conditions that are available out there, especially for model organisms under conditions of varied lighting and underwater conditions like turbidity. The work under consideration proves that UWGAN can have a number of positive outcomes, such as when it comes to color cast removal, and preserving color accuracy when light conditions are not uniform in underwater environments remains a problem due to its fluctuations in properties of light absorption and scattering that are special to the underwater environment scenes [54].

The CycleGAN framework first presented by Zhu et al.[32] presents unpaired image-to-image translation scenario which forms a strong baseline. In the paper, it comprises an architecture where each of the domains consists of a generator and a discriminator. Some of the features of the generator are residual blocks meant to represent more intricate mappings, this being a crucial feature for tasks that involve large-scale buildings, and feature-intense changes, including underwater-to-air image transformations. The used discriminator is the Patch GAN discriminator that works as follows: working on the small region of the image irrespective of the entire image. It improves the model's capacity to pay attention to small particulars and expands the model's means of development which leads to a reduction in computational resources used in the event of an occurrence. CycleGAN's success across a functionality that encompasses artistic style transfer and seasonality, is evident in other areas, confirming the versatility and variability of such image transformations as a useful technique for improving underwater image quality where paired data is not easy to get [32]. Together, these studies demonstrate CycleGAN's strengths in underwater image enhancement, enabling transformations that address color correction, noise reduction, and overall image clarity without requiring paired data. Despite the encouraging results, limitations remain, particularly with CycleGAN's sen-sitivity to domain variability and challenges in managing geometric transformations or mode collapse. Future research could aim to refine CycleGAN architectures or explore hybrid models to further enhance color fidelity, clarity, and adaptability in diverse underwater settings [53, 54, 32].

*3.4.3. ConditionalGAN*

Numerous studies have investigated GAN-based techniques for enhancing underwater images, addressing challenges like color distortion, reduced contrast, and texture degradation due to underwater light scattering and absorption. Notable among these are FUnIE-GAN, FE-GAN, and Sea-Pix-GAN which utilize conditional GAN frameworks to deliver high-quality underwater image enhancements across different operational conditions.

FUnIE-GAN, proposed by Islam et al.(2023)[34], is optimized for real-time image enhancement on single-board devices, making it particularly suitable for visually-guided underwater robotic applications. This model employs a simplified U-Net generator with reduced parameters to enable rapid inference while retaining essential image quality enhancements. The PatchGAN discriminator in FUnIE-GAN evaluates texture fidelity at the patch level, helping to preserve local details crucial for underwater image clarity. Evaluation metrics, including PSNR, SSIM, and UIQM, demonstrate FUnIE-GAN's effectiveness in enhancing perceptual quality through improved sharpness and color correction. However, its reliance on a paired dataset for training and its simplified architecture may reduce adaptability in diverse underwater environments with varying lighting and water turbidity.

FE-GAN, introduced by Han et al. (2023)[35], is slightly different from the models above in its implementation. It is another approach subjected to both productivity and innovation objectives. This makes it fit for real-time robotic applications in which real-time processing and low latency are a significant factor. The generator component integrated into the model is known as Hierarchical Attention Encoder. With this, they proposed the use of dense (HAE) with dual residual blocks (DRB), which allow multi-level feature extraction and its own memory requirements. This kind of architectural design helps FE-GAN to attain accurate computing, with results in PSNR, SSIM, and Mean Average Precision (mAP), which shows high performance. Despite its efficiency, FE-GAN relies upon synthetic datasets which might reduce its generalization of such naturally diverse scenes in underwater environments.

Sea-Pix-GAN, proposed by Chaurasia and Chhikara (2024)[36], approaches underwater image enhancement as an image-to-image translation task, focusing on achieving high-quality static image enhancement. It uses a U-Net-based encoder-decoder generator with skip connections to retain spatial details, and a PatchGAN discriminator to enforce local texture consistency, preserving the finer details in enhanced images. The loss function combines adversarial and L1 losses, helping to balance realistic output with color fidelity. Sea-Pix-GAN's effectiveness is validated using PSNR, SSIM, and UIQM metrics, which show notable improvements in image contrast and color balance. However, Sea-Pix-GAN sometimes faces challenges with high-resolution images, as the reliance on Euclidean loss can limit its ability to capture fine, high-frequency details, potentially introducing artifacts under certain conditions.

Across all three models[34][35][36], the use of conditional GANs is a common feature, allowing each model to generate realistic, high-quality underwater images conditioned on degraded inputs. The encoder-decoder generator architectures combined with PatchGAN discriminators provide enhanced texture and color restoration by focusing on pixel-level details. FUnIE-GAN stands out for its streamlined U-Net architecture suited for real-time applications, FE-GAN for its efficient hierarchical attention approach optimized for robotic tasks, and Sea-Pix-GAN for comprehensive image translation suited for static applications. This diversity in design illustrates the adaptability of GANs for specific underwater enhancement needs, from real-time, low-power solutions to high-quality static image enhancement.

Together, these studies demonstrate the capabilities of conditional GANs in underwater image enhancement, providing solutions that cater to various applications and performance requirements. While each model addresses core underwater imaging challenges, limitations in generalizability and resolution fidelity remain. Future research could aim to refine these architectures or explore hybrid models to further enhance performance and adaptability across diverse underwater conditions.

*3.4.4. DenseGAN*

In DenseGAN, proposed by Guo et al.[38] stimulates common challenges in underwater imagery: color shift, low exposure, and blurriness due to scattering and absorption of light in underwater environments. The methodology involves a generator network and a discriminator network. The generator integrates a Residual Multi-Scale Dense Block (RMSDB) within a fully convolutional structure, enabling it to generate enhanced underwater images. The discriminator aims to distinguish these created images from real underwater ones, helping the generator improve realism. Enhancing performance further, the model incorporates a deep residual learning framework to enhance optimization, also a dense block architecture that facilitates effective feature flow and reuse, boosting both accuracy and training effectiveness.

The generator architecture starts with two convolutional layers that progressively reduce the input feature map size. The first convolutional layer uses a $7 \times 7$ kernel and outputs 64 feature maps, followed by a second convolution with a $3 \times 3$ kernel that produces 128 feature maps. Both layers are followed by batch normalization, and the activation function Leaky ReLU, with a slope of 0.2, is applied. Each MSDB module merges three or four feature maps by concatenation, incorporating features from the previous layer to capture finer local details. It utilizes various kernel sizes ($1 \times 1$, $3 \times 3$, $5 \times 5$) to extract multi-scale features. The final $1 \times 1$ convolutional layer in each MSDB acts as a bottleneck, facilitating feature fusion and enhancing computational efficiency. After the RMSDB block, two deconvolution layers are used to reconstruct the image. The last deconvolution layer adjusts the output to match the input channel size, applying a Tanh activation function to ensure the output falls within the [1, 1] range, aligning with the original input distribution.

The discriminator network comprises five layers, incorporating spectral normalization to enhance training stability. The architecture follows a structure similar to the 70 × 70 PatchGAN used in previous works like pix2pix[36] and CycleGAN[32]. Notably, batch normalization (BN) to the first and last layers is not applied, but all intermediate convolutional layers follow the standard convolution-BN-LReLU design. Spectral normalization is used to control the Lipschitz constant of the discriminator, helping to maintain stable training. This technique is both computationally efficient and easy to implement.

During the training process, both the training set which consists of 6128 image pairs and the test set of 215 images have dimensions of 256 × 256 × 3 and are normalized. The Adam optimizer is used with a learning rate of 0.0001, and the batch size is set to 32. For each update of the generator, the discriminator is updated five times. The entire model is trained for 60 epochs. The network performs well on both synthetic and real underwater images. The model outperformed on state-of-the-art methods such as CycleGAN[32] and UGAN[53].

*3.4.5. MEvo-GAN*

Multi-scale Evolutionary Generative Adversarial Networks, or MEvo-GAN, is an innovative extension of traditional GANs that brings powerful new techniques for image enhancement. This framework introduces improvements across several areas: an optimized loss function, advanced deep residual shrinkage layers, and the use of multi-scale generative networks to capture minute details. Additionally, MEvo-GAN uses a genetic algorithm to improve training stability, choosing the best "descendant" network models at each step.MEvo-GAN offers two main benefits. First, it adopts a multi-path architecture for both its generator and discriminator, which enables it to capture an extensive array of features at different scales. This multi-scale processing is particularly beneficial for restoring the intricate details and textures of underwater images, where light and texture variations often present obstacles. Secondly, MEvo-GAN incorporates customized quality metrics specifically designed for underwater images which helps in refining the genetic algorithm to fine-tune the parameters. By drawing on principles from evolutionary biology such as selection, recombination, and adaptation MEvo-GAN achieves greater stability and enables faster convergence also optimizes the generator's parameters.

The architecture of MEvo-GAN consists of two main generators, labeled $G_{x \rightarrow y}$ and $G_{y \rightarrow x}$, alongside two discriminators, $D_x$ and $D_y$. The generator $G_{x \rightarrow y}$ is designed to transform low-quality underwater images into enhanced versions, while $G_{y \rightarrow x}$ handles the reverse transformation. The discriminators $D_x$ and $D_y$ evaluate the authenticity of the generated images, helping maintain high quality and realism in the outputs. In the initial setup, MEvo-GAN begins by creating a diverse group of generators, each assigned a unique set of parameters. Each of these ancestor generators then produces a series of 'descendants' by applying various mutation strategies. Following this, the offspring generators undergo evaluation, and the highest-performing ones are selected to become the parent generators for the next iteration.

In the training setup, the images are organized into two groups: low-quality underwater images were stored in the TrainA folder, while high-quality images were placed in the TrainB folder. To optimize the balance between speed and memory efficiency, the input images are resized to 256 × 256 pixels. The batch size is set to 1 and the training duration is set to 200 epochs. The evolutionary algorithm was designed to produce three offspring with each generation. To ensure the learning process balanced exploration and exploitation, the adaptive parameters are set to 1 and 0.1. For a more intuitive way to monitor the network's progress, the Visdom tool was utilized. It enabled us to save and visualize the reconstruction results every five iterations, giving a clear view of how the network was evolving over time. For parameter initialization, the Kaiming algorithm and Adam optimizer were used with initial learning rates of $1 \times 10^{-3}$ for the generator and $2 \times 10^{-3}$ for the discriminator, ensuring a balanced and efficient optimization process.

MEvo-GAN's experimental results were compared against several existing underwater image enhancement algorithms, demonstrating its superior performance across various datasets like EUVP[42], UIEB[40], and UFO 120[55]. In the color chart recovery test, MEvo-GAN excelled at restoring true colors, outperforming methods like WaterGAN[30], FunieGAN[34], and Shallow-UWnet, which either produced darker images, color intermingling, or a grayish tone. It also outperformed other methods in metrics like PSNR, SSIM, and UCIQE, particularly excelling in UCIQE, highlighting its advantage in overall image quality enhancement. This success is attributed to MEvo-GAN's multi-branch architecture, which includes sub-models like conv1, conv2, conv3, and conv4, each designed to capture texture, local features, contours, and color information at different scales. The integration of these features results in significantly enhanced image details, contrast, and color fidelity, making MEvo-GAN a powerful tool for underwater image enhancement.

**4. Loss functions**

Loss functions are a form of mathematical equations or metrics designed to measure the discrepancy that exists between the resultant output and the actual expected value. They are a tangible solution that guides the training process so the model can adjust them to lower the number of mistakes, therefore increasing precision. Below is a brief description of the reviewed loss functions in the research conducted.
**Adversarial Loss:** Adversarial loss plays an important role in conducting a competition between generator and Discriminator training to get better images of underwater scenes. This loss makes the generator produce images that mimic the target domain while enabling the discriminator to learn whether a given image is real or generated.

Table 2: A comparative table on various GAN methods

| Method | Publication Details | Main Idea and Methodology | Datasets | Loss Functions | Evaluation Metrics | Applications | Limitations or Challenges |
|---|---|---|---|---|---|---|---|
| WaterGAN[30] | Li et al., IEEE Robotics and Automation Letters, 2017 | Combines GAN and color correction networks. Generator: Three stages—Attenuation, Scattering, and Camera Model for underwater effects. Discriminator: Convolutional layers with Leaky ReLU activations. Color Correction Network: Encoder-decoder structure with skip connections for depth estimation and image restoration. | MHL test tank, Port Royal, Lizard Island | Euclidean loss | Euclidean distance, variance in RGB-space; RMSE for depth | Marine Biology, Marine archaeology, 3D Reconstruction | Assumes centered vignetting, difficulty with complex lighting patterns. |
| FW-GAN[52] | Wu et al., Signal Processing: Image Communication, 2022 | Multi-scale Fusion GAN (FW-GAN) for underwater image enhancement. Generator: U-Net-based architecture with multi-scale fusion connections and prior inputs (WB, CLAHE, DCP). Discriminator: PatchGAN with Spectral Normalization (SN) for stability and high-frequency detail preservation. | EUVP: Test-E500, UIEB: Test-U90, Test-C60, SQUID: est-S16 | Adversarial loss, L1 loss | Test-E500: SSIM: 86%, PSNR: 27.18 dB. Test-U90: SSIM: 92%, PSNR: 26.65 dB, Entropy: 7.69, Twice Mixing: 2.68. Test-C60: Perception Score (PS): 2.93, UCIQE: 0.598, UIQM: 2.84, Entropy: 7.48, Twice Mixing: 1.49. Test-S16: Perception Score (PS): 3.28, UCIQE: 0.593, UIQM: 2.47, Entropy: 7.68, Twice Mixing: 1.45 | Improves underwater vision for robots; Feature detection and saliency prediction | High computational load; Reliance on Prior Features. |
| UGAN[53] | Fabbri et al., Presented at IEEE ICRA, 2018 | UGAN improves underwater image quality reconstructing distorted underwater images. Dataset Generation: CycleGAN, GAN Architecture: Generator: U-Net-based encoder-decoder architecture with skip connections. Discriminator: PatchGAN with WGAN-GP. | ImageNet subsets and Flickr images for training and testing. Generated synthetic pairs with CycleGAN. | Wasserstein loss, L1 loss, Gradient Difference Loss. | Edge detection (Canny), Gradient Difference Loss (GDL), Mean and Standard deviation, and Diver Tracking Algorithm. | Enhances visual tasks in underwater robotics, diver tracking, and marine exploration | Limited real underwater datasets, complexity in capturing natural distortions, potential blur in output. |
| UWGAN[54] | Li et al., IEEE Signal Processing Letters, 2018 | Uses a weakly supervised model with GANs. Generator: CycleGAN, Discriminator: 70×70 PatchGAN. Loss Functions: Adversarial Loss, Cycle consistency loss, and SSIM loss. | 3800 underwater images and 3800 air images | Adversarial Loss, Cycle consistency loss. | Visual quality assessment, User study, Saliency detection, Keypoint matching | Enhancing underwater visual tasks like saliency detection and keypoint matching | Color fidelity is challenging due to unpaired datasets and variable underwater conditions. |
| CycleGAN[32] | Zhu et al., IEEE International Conference on Computer Vision (ICCV), 2017 | Proposes CycleGAN for unpaired translation using a cycle consistency loss. Generator: G (X → Y) and F (Y → X) translate images between domains. Discriminator: 70×70 PatchGAN | Diverse domains, unpaired collections from ImageNet, Flickr, etc. | Adversarial Loss, Cycle consistency loss. | Perceptual Studies (Amazon Mechanical Turk), FCN Score and Semantic Segmentation Metrics | Object transfiguration, Season transfer, Artistic style transfer, and Photo enhancement. | Geometric changes; challenges with mappings and dataset specificity. |
| FUnIE-GAN[34] | Islam et al., IEEE Robotics and Automation Letters, 2020 | A conditional GAN model designed for real-time underwater image enhancement, focusing on color correction and contrast adjustment. Generator: Simplified U-Net encoder-decoder, Discriminator: PatchGAN | EUVP | Conditional Adversarial Loss, L1 loss, Cycle-Consistency Loss | PSNR- 21.92, SSIM- 0.8876 | Real-time underwater applications, object detection | May lack robustness in varied underwater settings; Training limited to paired data |
| FE-GAN[35] | Han et al., Electronics, 2023 | AFE-GAN is a fast and efficient conditional GAN model. Generator: Hierarchical Attention Encoder with Dual Residual Blocks(DRB), Discriminator: Markov discriminator (PatchGAN). | EUVP, ImageNet, Mixed datasets(NYU-v2 and UIEBD) | Adversarial loss, L1 loss, and perceptual loss | EUVP: PSNR- 26.83, SSIM- 0.8779, UIQM- 2.87. ImageNet: PSNR- 24.5396, SSIM- 0.8106. Test-R90 PSNR-20.68. Test-C60 UIQM-1.01 | Underwater Robotics, Ocean Exploration, Surveillance and Monitoring. | Limited generalization to diverse underwater conditions; heavily reliant on synthetic datasets for training |
| Sea-Pix-GAN[36] | Chaurasia et al., Journal of Visual Communication and Image Representation, 2024 | Sea-Pix-GAN enhances underwater images by addressing color distortion, noise, and contrast issues. Generator: Encoder-decoder with skip connections (U-Net-based), Discriminator: PatchGAN. | EUVP Dataset | Conditional Adversarial Loss, L1 loss. | PSNR- 23.30, SSIM- 0.79, UIQM- 2.84 | Underwater image enhancement, marine research | Struggles with high-resolution image inputs |
| DenseGAN[38] | Guo et al., IEEE Journal of Ocean Engineering, 2020 | Generator: uses RMSDB and employs residual learning final output mapped using a Tanh activation function. Discriminator: PatchGAN and Spectral normalization. | 6,128 paired images generated via CycleGAN, 215 underwater images from research and public datasets. | L1 loss, Gradient loss. | UCIQE: 0.614, UIQM: 3.888 for real-world underwater images, UCIQE: 0.628, UIQM: 3.954 for synthetic underwater images | Marine exploration, robotic vision, and oceanography. | Artifacts in Synthetic Data due to aesthetic issues |
| MEvo-GAN[39] | Fu et al., Journal of Marine Science and Engineering, 2024 | Generator: Multi-path with dilated convolutions and residual blocks for detail enhancement and noise reduction. Discriminator: Multi-scale structure for realism. | EUVP, UIEB, and UFO-120 | Adversarial loss, Cycle-consistency loss, Identity-consistency loss, and Perceptual loss. | PSNR-21.2758, SSIM- 0.8662 and UCIQ-0.6597 | Marine resource management, ecological monitoring, and ocean exploration | Faces challenges with realistic synthetic images and high training costs |

In the generator, the adversarial loss function can be written as:

$$L_{adv}(G, D_{U_e}) = E_{U \sim p_{data}(U)} \left[ \log D_{U_e}(G(U)) \right] \quad (1)$$

Where G(U) is the output of an image from a generator with respect to the underwater image input U and where $D_{U_e}$ discriminator for finding out the probability of whether generated images are real or fake. The loss function for the discriminator thus consists of two terms— one on real images and one on generated images which can be written as:

$$L_{adv}(D_{U_e}, G) = -E_{e \sim p_{data}(e)}\left[\log D_{U_e}(e)\right] - E_{U \sim p_{data}(U)}\left[\log(1 - D_{U_e}(G(U)))\right] \quad (2)$$

e is a real image sampled from the target distribution. This loss function is maximized by the discriminator, but minimized by the generator, as it generates images that seem more and more realistic[39, 56].

**L1 Loss:** Mean Absolute Error (MAE) loss, or L1 loss, is used in image generation to capture fine details and avoid blurring by minimizing the average absolute difference between a generated image $G(U, W)$ and a target image $V$[57]. Mathematically, it is expressed as:

$$L_1(G) = E_{U,V,W}[\|V - G(U, W)\|_1] \quad (3)$$

where $V$ is the reference image, and $G(U, W)$ is the generated output. By minimizing this loss, the model produces sharper images that align closely with the target, improving visual quality.

**L2 Loss:** The L2 loss, which is also called the Euclidean loss is a commonly used loss function in conditions that call for compensating the distance between predicted and actual values, in regression tasks or on image generation. It operates by taking for every single value, the square of the difference between the predicted and actual value[30].

$$L_2 = \sum_{i=1}^{X} (U_i - \hat{U}_i)^2 \quad (4)$$

**Perpetual Loss:** Perceptual loss is a technique that encourages generated images to resemble real images more closely by comparing their high-level feature representations rather than just individual pixel values. This is typically achieved using a pre-trained neural network, like VGG-19, which has been trained on extensive datasets and is effective at capturing intricate visual details. Given a generated image $G(x)$ and a target image $y$, both are fed into the VGG-19 network, and feature maps $\Phi_j(G(x))$ and $\Phi_j(y)$ are extracted from specific layers $j$. The perceptual loss[58], denoted as $L_{per}$, is calculated as the L1 norm of the difference between these feature maps, formulated as: The perceptual loss, $L_{per}$, is defined as:

$$L_{Perceptual} = \sum_{m=1}^{M} \alpha_m \|\psi_m(F(z)) - \psi_m(T)\|_1 \quad (5)$$

where $m$ is the index of the feature layers $\alpha_m$ is the weight associated with the $m$-th feature layer, controlling its contribution to the total loss, $\psi_m$ is the feature extractor at layer $m$, $F(z)$ is the generated or enhanced image produced by the model from the input $z$, $T$ is the target image, $\|\cdot\|_1$ denotes the L1 norm, which calculates the absolute difference between the feature maps of the generated image $F(z)$ and the target image $T$ at layer $m$, $M$ is the total number of layers considered for the perceptual loss computation.

**Cycle consistency Loss:** Cycle consistency loss helps ensure that when an image is transformed from one domain to another and then back again, it retains its original appearance[59]. This loss encourages the model to learn an accurate mapping between the source and target domains, helping to prevent mode collapse (where the model generates similar outputs regardless of input variations). Cycle consistency loss has two components: one for transforming from source to target and back, and another for the reverse transformation. The cycle consistency loss, $L_{cyc}(A, B)$, is defined as:

$$L_{cycle}(A, B) = E_{x \sim P_{data}(x)}[\|B(A(x)) - x\|_1] + E_{y \sim P_{data}(y)}[\|A(B(y)) - y\|_1] \quad (6)$$

where, $A$ and $B$ are generators that map images between two domains, $x \sim P_{data}(x)$ and $y \sim P_{data}(y)$ represent samples from the data distributions of the two domains, $\|\cdot\|_1$ is the L1 norm, measuring the absolute difference between the original and reconstructed images.

**Wasserstein Loss:** Wasserstein loss is part of Wasserstein GANs (WGANs), which was developed to work in an attempt to enhance the stability in, as well as the efficiency of, GANs through the measurement of differences between the real and generated data based on the Earth Mover's (Wasserstein) distance. WGAN does not output a binary decision of "is this real" or "is this fake," but generates real values instead. It involves 'teaching' the critic to make the scores of real samples as far from scores of generated samples as possible; on the other side, the generator is 'teaching' the critic to reduce this difference as much as possible to produce higher quality data. To ensure that the critic has the correct form of 1-Lipschitz, most notably WGAN-GP incorporates a gradient penalty. The Wasserstein loss, denoted as

$$L_{WGAN}(G, D) = E_{a \sim P_{data}}[D(a)] - E_{b \sim P_b}[D(G(b))] + \lambda_{GP} E_{\tilde{a} \sim P_{\tilde{a}}}\left[(\|\nabla_{\tilde{a}} D(\tilde{a})\|_2 - 1)^2\right] \quad (7)$$

where **a** represents real data samples from $P_{data}$, and **b** represents noise samples from the prior distribution $P_b$. The generator $G(b)$ takes the noise **b** as input and produces a generated output, while $D(\cdot)$ is the critic that assigns scores to both real and generated samples. Interpolated samples between real and generated data are denoted as $\tilde{a}$, drawn from $P_{\tilde{a}}$, and $\|\nabla_a D(\tilde{a})\|_2$ represents the $L_2$-norm of the gradient of $D$ with respect to these interpolated samples. The term $\lambda_{GP}$ controls the strength of the gradient penalty, ensuring that the critic satisfies the Lipschitz constraint[53].

**Conditional Adversarial Loss:** The conditional adversarial loss most commonly applied to cGANs promotes the generation of realistic data for a given condition $X$. It is an integral part of image-to-image mapping, the goal of which is to obtain the result $G(X, Z)$ that is as close as possible to the real data $Y$, given the condition $X$. It is expressed as:

$$L_{cGAN}(G, D) = E_{X,Y}[\log D(Y)] + E_{X,Y}[\log(1 - D(X, G(X, Z)))] \quad (8)$$

where $D$ represents the discriminator with distinguishing real samples $Y$ from generated ones $G(X, Z)$, and $Z$ is a random noise vector. The generator $G$ learns to produce outputs $G(X, Z)$ that maximize $\log D(X, G(X, Z))$, pushing $D$ to misclassify them as real.

By minimizing this loss, the model achieves both realism in the generated data and alignment with the input condition, making it highly effective for tasks requiring context-aware generation[34][36].

**Identity consistency Loss:** Identity consistency loss encourages the generator to preserve essential features of an input image, even after transformation, by keeping the generated output as close to the original as possible. This helps in reducing any unwanted changes or information loss during the transformation process. The identity consistency loss, $L_{id}(A, B)$, is defined as:

$$L_{id}(A, B) = E_{y \sim P_{data}(y)}[\|A(y) - y\|_1] + E_{x \sim P_{data}(x)}[\|B(x) - x\|_1] \quad (9)$$

where, $A$ and $B$ represent the transformations between two domains, $y \sim P_{data}(y)$ and $x \sim P_{data}(x)$ denote samples from the data distributions of the two domains, $\|\cdot\|_1$ is the L1 norm, which measures the absolute difference between the input and the generated output. By minimizing this loss, the model is guided to keep the core details and appearance of the input image intact, resulting in more faithful transformations[39].

**Gradient Loss:** Gradient loss is a type of loss function that targets discrepancies in edge strength (gradients) between the predicted and actual images[60]. Unlike standard pixel-based loss functions like L1 or L2, which evaluate pixel-by-pixel differences, gradient loss emphasizes the image's edges and transitions. By doing so, it helps retain and sharpen fine details. This makes gradient loss particularly valuable for tasks such as underwater image enhancement, where water conditions often blur edges. Restoring these edges not only improves the image's visual quality but also aids in better object detection and overall scene clarity. The gradients in the $x$- and $y$-directions for an image $U$ are computed as:

$$\text{Gradient}_x(U) = U(i, j) - U(i + 1, j)$$

$$\text{Gradient}_y(U) = U(i, j) - U(i, j + 1)$$

Where $U(i, j)$ represents the pixel intensity at location $(i, j)$. The gradient loss between an enhanced image $U_e$ and an original image $U$ is expressed as:

$$L_{gradient} = \frac{1}{n} \sum_{i,j} \left( |\text{Gradient}_x(U) - \text{Gradient}_x(U_e)| + |\text{Gradient}_y(U) - \text{Gradient}_y(U_e)| \right) \quad (10)$$

where, $n$ is the total number of pixels, $\text{Gradient}_x$ and $\text{Gradient}_y$ are the gradients in the $x$- and $y$-directions, respectively.

## 5. Limitations and Challenges

There are several challenges that are inherent in enhancing underwater images using GANs due to the conditions that are inherent when capturing underwater images. Some models rely on centered vignetting where the image becomes darker or softer as you move away from the center. However, this assumption is ineffective for real underwater scenes, where lighting is irregular and unpredictable which results in inconsistent outcomes when such models are applied to practical problems. Further, the high degree of color scattering and turbidity, especially underwater environment demands high computational power, which makes the applicability and possibility of implementing such models in regular hardware a challenging proposition. There is a limited availability of diverse real underwater datasets that limits the training of models, affects generalization and when tested on unknown data, provides blurry or wrong images.

Color fidelity is still an issue of concern because other underwater conditions such as depth and visibility can cause extreme color distortion which hampers models trained on synthetic or unpaired data. Additionally, many underwater images are distorted by the movement of the water and water refraction, which is problematic for GANs as they plateau or remain unvaried. Finally, high computational costs and the usage of synthetic data to train GANs that may not fully resemble the true underwater environment limit the applicability of GANs in real-life underwater image enhancement tasks.

## 6. Conclusion

This paper presents a review of GAN-based methods for enhancing images from an underwater environment, in addition to discussing advancements over traditional and CNN-based methods. WaterGAN, CycleGAN, ConditionalGAN, DenseGAN, and MEvo-GAN have been identified as essential GAN architectures for enhancing underwater imaging by overcoming color distortion, low contrast, and scattering effects on real-world underwater applications ranging from marine research to underwater robotics. We reviewed datasets used for training and testing, Evaluation metrics, and Loss functions assessing their applicability and reliability in various underwater environments.

However, several challenges are still associated with GAN-based underwater image enhancement though there has been considerable improvement made. Present models are trained using synthetic or unpaired underwater datasets which might not be always relevant to different underwater scenarios, consume much computational power, and can not provide precise color restoration and image clarity in different conditions. For future research works, it is necessary to explore the improvements of the GAN architectures to suit the real-time applications and work on the limitations of the datasets to make them more generalized to the different underwater environments.

**Competing Interests**

The authors declare that they have no competing interests.